\documentstyle[twocolumn,prb,aps,psfig,amssymb]{revtex}
\begin{document}
 \twocolumn[\hsize\textwidth\columnwidth\hsize\csname @twocolumnfalse\endcsname
\draft
\begin{title}
{ Parity Effects in Stacked Nanoscopic Quantum Rings }
\end{title}
\author{ Kang-Hun Ahn and Peter Fulde}
\address{Max-Planck-Institut f\"ur Physik Komplexer Systeme, 
N\"othnitzer Strasse 38, 01187 Dresden, Germany      \\ }
\date{\today}
\maketitle
  
\widetext

\begin{abstract}
The ground state and the dielectric response of stacked
quantum rings are investigated in the presence of an applied magnetic
field along the ring axis.
For odd number $N$ of rings and an electric field perpendicular to
the axis, a linear Stark effect occurs at distinct
values of the magnetic field.
At those fields energy levels cross in the absence of electric field.
For even values of $N$ a quadratic Stark effect is expected in all cases, 
but the induced electric polarization is discontinuous 
at those special magnetic fields. 
Experimental consequences for related nanostructures are discussed. 
\vspace{0.5cm}
\end{abstract}
]
\narrowtext


Considerable progress has been made in precise measurements of
the dielectric response of solids at low temperatures.
New physical effects have been found due to these advances.
An example are certain multicomponent glasses where it has been found that
at ultra-low temperatures T $<$ 100mK the dielectric response is
strongly magnetic field dependent\cite{strehlow}.
The interactions between the different tunneling systems in the glasses
consisting of groups of atoms ( or single atoms) 
are sufficiently strong so that large groups of them form a coherent state with
a Aharonov-Bohm phase\cite{strehlow,kettemann}.
Here we are considering a stack of symmetrical ring molecules
(quantum rings) with one (unpaired) electron per ring.
Electrons in different rings interact with each other.
For simplicity we restrict that interaction to neighboring rings, 
but the conclusions of the present investigation
do not depend on this simplification.
The purpose of this communication is to point out 
that a stack of such  quantum rings shows an interesting
even-odd effect with respect to the number $N$ of rings in the stack.
In particular we want to show the following.

(i) The dielectric response of an even number of rings to an electric
field in the plane of the rings is considerably smaller than in the 
case of an odd number.

(ii) The induced electric polarization $P(\phi)$ as a function of
magnetic flux $\phi$ through the rings is a continuous function
of $\phi$ when the number of rings is odd while it shows discontinuities
in the case when $N$ is even.

 In order to prove the above statements we start from a Hamiltonian
for the cofacially stacked rings in an axial magnetic field and
a perpendicular electric field 
with simplified electron interactions.
The interactions are assumed to take place only between neighboring 
rings and we assume for them a simple angular dependence.
In analogy to previous works on glasses\cite{kettemann}, 
we write for a system of $N$ rings
\begin{eqnarray}
H&=&H_{0}-{\mathcal E}p_{0}\sum_{j}\cos\theta_{j},\\
H_{0}&=&\frac{\hbar^{2}}{2mr^{2}}\sum_{j}
\left(i\frac{\partial}{\partial \theta_{j}}+\frac{\phi}{\phi_{0}}\right)^{2}
+J\sum_{j}^{N-1}\cos(\theta_{j+1}-\theta_{j}).
\label{hamiltonian}
\end{eqnarray}
Here $r$ is the radius of a ring, $m$ is the particle mass,
$\phi_{0}=hc/e$ is the flux quantum, $J>0$ is the coupling
strength  of particles in neighboring rings, ${\mathcal E}$ is
the size of the electric field in  the plane of the rings and
$p_{0}= er$ is a dipole moment.
The following considerations do not depend on the special form of
the interaction chosen here provided the interaction is repulsive.
In order to simplify the notation we shall use the abbreviation
$u=\hbar^{2}(2mr^{2})^{-1}$.
First we consider the case of ${\mathcal E}=0$.
Then the total angular momentum commutes with $H_{0}$ and hence is 
a good quantum number.
Disregarding the spin indices
which are
 not important in this context,
the eigenfunction of $H_{0}$ 
 can be classified by two
indices $M$, i.e., the azimuthal total angular momentum quantum number, 
and $\alpha$.
The quantum number $\alpha$ describes the different states
 of a given angular 
momentum $M$.
In the absence of any interactions, i.e., for $J=0$ the eigenstates
are highly degenerate, but this degeneracy is lifted by the interactions.
It is easy to show from Eq. (\ref{hamiltonian}) that in the presence of
 a flux $\phi$ through the rings the eigenenergies are given by
\begin{eqnarray}
E_{M,\alpha}(\phi)=
E_{M,\alpha}(0)+uN\left(\frac{\phi}{\phi_{0}}-\frac{M}{N}
 \right)^{2}-u\frac{M^{2}}{N},
\label{e_rel}
\end{eqnarray}
where
$E_{M,\alpha}(0)$  are the eigenenergies  in the absence of an applied
magnetic field.
As a function of $\phi$ the eigenenergy $E_{M,\alpha}(\phi)$ and
in particular the ground-state energy $E_{M,G}(\phi)$ have
the form of a parabola centered at $\phi/\phi_{0}=M/N$.
As for the ground state we need to consider therefore 
$M$ values only within the range $0\leq M \leq N$ when $\phi$ varies
between $0\leq \phi \leq \phi_{0}$.
Because of the $\phi$ dependent term in Eq. (\ref{e_rel}), 
energy levels  cross each other at fluxes
\begin{eqnarray}
\phi_{M,M^{\prime}}^{\alpha,\alpha^{\prime}}
=\frac{\phi_{0}}{2u(M-M^{\prime})} 
\left[ E_{M,\alpha}(0)-E_{M^{\prime},\alpha^{\prime}}(0)\right].
\label{crossvalue}
\end{eqnarray}
When we consider the ground states ($\alpha=G$), 
then with increasing flux a cross-over will take place from
the state with $E_{M,G}(\phi)$ to the one with $E_{M+1,G}(\phi)$,
starting from $E_{0,G}(\phi)$.
This is a conjecture at this point which is equivalent to the assumption that
\begin{eqnarray}
\frac{\partial^{2} E_{M,G}(0) }{\partial M^{2}}
= E_{M+1,G}(0)+E_{M-1,G}(0)-2E_{M,G}(0) > 0.
\label{conjecture}
\end{eqnarray}
We do not provide a proof here of that inequality but merely
mention that in the case of strong interactions, i.e., for 
 $J\rightarrow \infty$
this is immediately seen.
The system behaves like a single quantum ring with a particle
of mass $Nm$ and a charge $Ne$. The ground-state energy is
\begin{eqnarray}
E_{M,G} = \frac{\hbar^{2}}{2Nmr^{2}}\left(
N\frac{\phi}{\phi_{0}} -M\right)^{2}+ \cdots,
\label{e0}
\end{eqnarray}
where the remaining terms neither depend on $M$ nor on $\phi$.
From Eqs. (\ref{crossvalue},\ref{conjecture}) we obtain a ground state
with $N$ level crossing points
$\phi(1) < \phi(2) < \cdots < \phi(N)$.
At these points the ground state is degenerate.

\begin{figure}
\centerline{\psfig{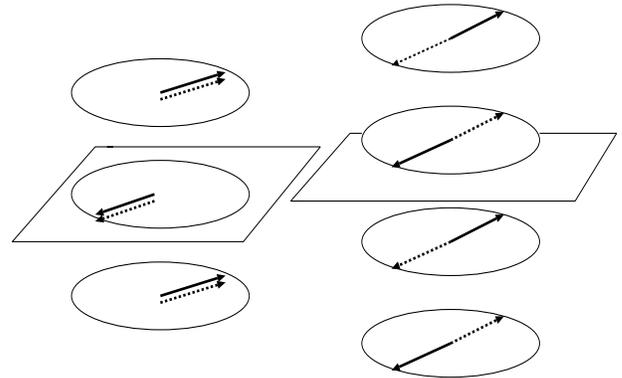}}
\caption{  A schematic drawing to explain the operation
 ${\mathcal P}$ on odd (left)
and even (right) number of rings.
The mirror is located in the middle of the each stack.
The dotted arrows are the mirror images of solid arrows.
Note that while for odd $N$ the mirror-reflected arrows coincide with the
original ones,
for even $N$ the arrows are rotated by $\pi$.  }
\label{parity}
\end{figure}

When an electric field ${\mathcal E}$ is turned on, the degeneracies
are lifted provided the corresponding matrix element
$\big<\Psi_{M+1,G} \big|\sum_{j}\cos\theta_{j} \big|\Psi_{M,G} \big>$ 
differs from zero.
We want to show that this is indeed the case when $N$ is odd while
when $N$ is even the matrix element vanishes.
For this purpose we introduce an reflection operator 
( with respect to the $xy$-plane in the middle of the stack of rings)
${\mathcal P}$ and a rotational operator $R_{\pi}$.
They are defined through 
\begin{eqnarray}
&& {\mathcal P}
\Psi (\theta_{1}, \theta_{2},\cdots,\theta_{N})
= \Psi (\theta_{N}, \theta_{N-1},\cdots,\theta_{1}),
\\&& R_{\pi}
\Psi (\theta_{1}, \theta_{2},\cdots,\theta_{N})
= \Psi (\theta_{1}+\pi, \theta_{2}+\pi,\cdots,\theta_{N}+\pi).
\end{eqnarray}
Note  that $\big[ {\mathcal P},H_{0}\big] 
=\big[ R_{\pi},H_{0}\big]=0$  and furthermore that
$\big[ {\mathcal P}, R_{\pi} \big]=0$.
Both ${\mathcal P}$ and $R_{\pi}{\mathcal P}$ have eigenvalues $\pm 1$.
Because of the repulsive interactions
the probability of finding particles in neighboring  rings
by an angle $\pi$ apart exceeds the case of finding these
with the same angles. This implies that
\begin{eqnarray}
{\mathcal P}\Psi_{M,G}= \Psi_{M,G}~~(N=odd)
\label{operation_odd}
\\
R_{\pi}{\mathcal P}\Psi_{M,G}= \Psi_{M,G}~~(N=even),
\label{operation_even}
\end{eqnarray}
because otherwise we have the unphysical condition
$\Psi_{M,G}(\theta,\theta+\pi,\theta,\cdots)=0$.
For an illustration see Fig.\ref{parity}.
When $N$ is even we may also write
${\mathcal P}\Psi_{M,G}=R_{\pi}\Psi_{M,G}$ and furthermore
$R_{\pi}\Psi_{M,G} =e^{i\pi M}\Psi_{M,G} =(-1)^{M}\Psi_{M,G}$.
Therefore, by inserting ${\mathcal P}^{2}=1$ into
the matrix element 
$\big< \Psi_{M+1,G} \big| \sum_{j}\cos\theta_{j}\big|\Psi_{M,G}\big>$ 
we find for these $N$ values
\begin{eqnarray}
\nonumber
&& \big< \Psi_{M+1,G} \big|
{\mathcal P}^{2}
 \sum_{j}\cos\theta_{j}
{\mathcal P}^{2}
\big| \Psi_{M,G} \big>
\\&& =
(-1)^{2M+1}\big< \Psi_{M+1,G} \big| \sum_{j}\cos\theta_{j}
\big| \Psi_{M,G} \big>=0.
\label{offdiagonal}
\end{eqnarray}
This implies that for $N$=even  the ground state remains degenerate
at the crossing points of fluxes $\phi(1),\cdots,\phi(N)$ even when an 
electric field is applied in the plane of the rings.

Next we consider the electric polarization $P(\phi)$ which is
defined by $P(\phi)=p_{0}\sum_{j}
\big< \Psi_{G}(\phi)\big|\cos\theta_{j} \big|\Psi_{G}(\phi)\big>$, 
where $ \big|\Psi_{G}(\phi)\big>$ is the ground state 
in the presence of the magnetic flux $\phi$.
In the more general case, a thermodynamic average has to be taken instead.
At flux $\phi$ away from the special cases $\phi(1),\cdots,\phi(N)$
an electric field causes a quadratic Stark effect.
But in the vicinity of $\phi(1),\cdots,\phi(N)$ a linear Stark effect
is expected, provided the electric field ${\mathcal E}$ splits
the degenerate states as is the case for odd values of $N$. 
Therefore the polarization is quite different in the two cases.

In the regime of a quadratic Stark effect we can apply non-degenerate 
second-order perturbation theory. 
For small values of $\phi$ the ground-state energy shifts
due to the electric field is
\begin{eqnarray}
\Delta E& = &\left( {\mathcal E} p_{0} \right)^{2}
\sum_{M= \pm 1} \sum_{\alpha }
\frac{\left|\left< \Psi_{M,\alpha} 
\left| \sum_{j}\cos\theta_{j}\right| \Psi_{0,G}\right> \right|^{2}}
{E_{0,G}(\phi)-E_{M,\alpha}(\phi)}.
\label{pertubation}
\end{eqnarray}
We abbreviate the matrix element 
$W_{\alpha}=\big| \big<  \Psi_{1,\alpha} \big| 
\sum_{j}e^{i\theta_{j}}  \big| \Psi_{0,G} \big> \big|^{2}$ 
and set $\Delta_{\alpha}=E_{1,\alpha}(\phi)-E_{1,G}(\phi)$.
When $N$ is odd the leading contribution comes from $\alpha=G$ 
with $\Delta_{\alpha=G}=0$.
In contrast, when $N$ is even we have $W_{G}=0$ and the
leading contribution has $\Delta_{\alpha \neq G} \propto J$ and therefore
is smaller.
This is particularly the case in the strong interaction limit where
\begin{eqnarray}
P_{odd}(\phi) &\approx& \frac{{\mathcal E}p_{0}W_{G}N}{u}
\left(1+4N^{2}\left(\frac{\phi}{\phi_{0}}\right)^{2} \right)
;~(\phi<<\phi_{0}/N),
\\
P_{even}(\phi) &\rightarrow& 0~ {\rm for} ~ J \rightarrow \infty.
\label{poddest1}
\end{eqnarray} 
In the region of a linear Stark effect which is
present only when $N$ is odd, we find from degenerate perturbation theory
$\Delta E \approx -\frac{1}{2}{\mathcal E}p_{0}\sqrt{W_{G}}$ and therefore 
\begin{eqnarray}
P_{odd}(\phi) \approx \frac{1}{2}p_{0}\sqrt{W_{G}}.
\label{poddest2}
\end{eqnarray}

For odd $N$,  
the ground state in the presence of the electric field
is non-degenerate for any $\phi$ and therefore
the corresponding wavefunction and the polarization $P(\phi)$
are also continuous functions of $\phi$. 
This is not the case for even $N$. 
Since level crossings at fluxes $\phi(1),\cdots,\phi(N)$ are not
removed by an applied electric field, $P(\phi)$ is generally
discontinuous at these special values of the magnetic field. 

\begin{figure}
\centerline{\psfig{figure=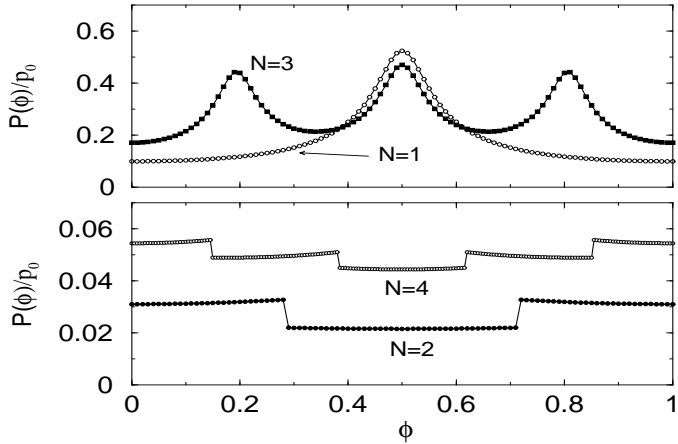,height=6cm,width=0.5\textwidth}}
\caption{ The induced polarization $P(\phi)$ in units of $p_{0}=er$
as a function of the threaded magnetic flux per ring $\phi$
 calculated by numerical diagonalization.
The interaction strength $J$ and the electric field ${\mathcal E}$ 
were chosen to be $J=2u$ and ${\mathcal E}p_{0}=0.1u$.  }
\label{POL}
\end{figure}

Shown in Fig. \ref{POL} are numerical results for 
$P(\phi)$ of rings up to $N=4$.
Calculations have been done by diagonalization of the respective 
Hamiltonian which is a matrix of size up to 3000$\times$3000.
The basis which is used in the calculations consists of 
products of $N$ single-particle eigenfunctions of the single-particle
Schr\"odinger equation in the presence of an electric field.
For particular values $J/u=2$ and ${\mathcal E}p_{0}/u=0.1$  
we find that $P_{even}$ is by about one order of
magnitude smaller than $P_{odd}$.
For $N=3$ a quasiperiodic peak structure is observed.
Note the discontinuities of the polarization for $N=2,4$ in 
accordance with the behaviour pointed out above.

Not only the dielectric response but also far-infrared spectroscopy
may show even-odd ring number parity effects.
For example, for stacks with an even number of rings the optical 
transition between the ground- and first excited state 
vanishes as seen from Eq.(\ref{offdiagonal}) while
for odd number of rings it does not.
Nanoscopic quantum rings can be realized on the surface of semiconductors
by using self-assembling techniques\cite{lorke}.
It seems possible that these method can be also used to build up
stacks of several rings. 
Other candidates are stacks of large ring molecules\cite{marks}.
However, those rings are not perfectly circular and also 
the inter-ring interactions are not limited to nearest
neighbors only.
For that reason it is useful to consider the more general
Hamiltonian\cite{kettemann}; 
\begin{eqnarray}
\nonumber
H_{0}&=&\frac{\hbar^{2}}{2mr^{2}}\sum_{j=1}^{N}
\left( i\frac{\partial}{\partial \theta_{j}}
+\frac{\phi}{\phi_{0}}  \right)^{2}
+\sum_{i<j}U_{|i-j|}(\theta_{i}-\theta_{j})
\\
&+&\sum_{j}V_{j}(\theta_{j}),
\label{gen_hamilt}
\end{eqnarray}
where 
$U_{|i-j|}(\theta_{i}-\theta_{j})$ is a repulsive 
particle interaction and
$V_{j}(\theta_{j})$ is the azimuthal barrier potential of the $j$-th ring.
When we consider only the (unpaired) electron in the highest
occupied molecular orbital, $V(\theta)$ is the potential
felt by it.
The extension of the range of the inter-ring interaction
$U_{|i-j|}(\theta_{i}-\theta_{j})$  does not affect the invariance of
it under the operation ${\mathcal P}$.
 
\begin{figure}
\centerline{\psfig{figure=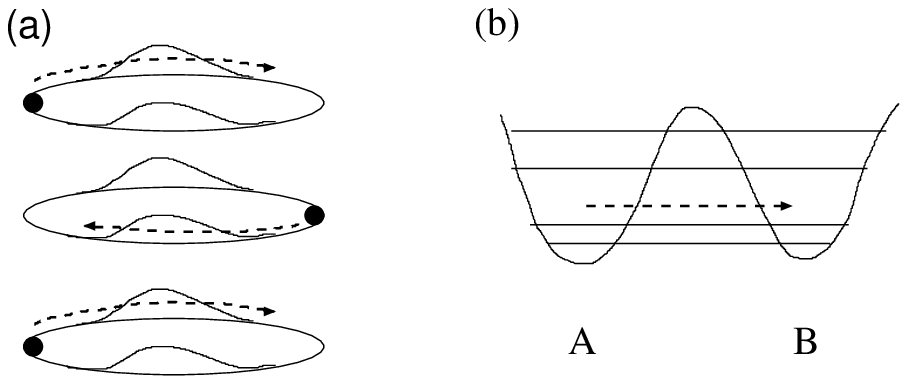,height=5cm,width=0.4\textwidth}}
\caption{ (a) A schematic drawing to illustrate the correlated tunneling
process. The dotted arrow represents a correlated tunneling path. 
In the case of correlated tunneling, the tunneling particles in neighboring
 rings are in opposite sides in order to reduce the effective barrier
enhanced by strong inter-particle interactions.
(b) A schematic drawing for the effective potential in the configuration
space (solid curve) and the splitted energy levels (solid lines)
 due to the correlated tunneling.  The dotted arrow represents the path
corresponding to the path in (a).  }
\label{tls}
\end{figure}

In order for $H_{0}$ to be invariant under the 
operation of ${\mathcal P}$, it must hold that $V(\theta)=V_{j}(\theta)$.
This is the case when the ring molecules are stacked without rotations
with respect to each other, i.e., when
$V(\theta)=V_{j}(\theta)$ for all $j$.
Level crossings will exist if $V(\theta)$ satisfy the condition
 $\int_{0}^{2\pi}d\theta V(\theta)\exp(i\theta)=0$.
An example is a symmetric double-well potential.
When the inter-ring interactions are larger than the potential
barriers in $V(\theta)$ and furthermore larger than the
kinetic energy $u$, the ground state is a superposition of
two configurations.
One of it is indicated in Fig. \ref{tls} (a) and the other one is
obtained by rotating all particles by $\pi$.
In the two configurations labeled 
$\big|A\big>$ and  $\big|B\big>$
the repulsive energy is minimized. 
Due to the kinetic energy both configurations are connected by
a tunneling matrix element resulting in correlated tunneling 
paths of the particles ( see Fig. \ref{tls} (a) ).
Within this classical path picture the two lowest lying
eigenstates can be looked upon as those of a single
quantum ring with an effective double well potential
and effective charge ( see Fig.\ref{tls} (b) ).
The flux dependent tunneling splitting is given by
$t_{\rm eff}(\phi)=t_{\rm eff}(0)\cos(N\pi\phi/\phi_{0})$.
One notices that the flux $N\phi$ enters here because
the correlated motion of the $N$ particles corresponds to
an effective charge $Ne$ of the system.
Level crossing takes place whenever $t_{\rm eff}(\phi)=0$.
For odd values of $N$ one notices that 
$\big| A\big>$ and $\big| B\big>$
are reproduced when ${\mathcal P}$ is acting on them.
The same holds true  for the states 
$\big| A\big> \pm \big| B\big>$.
In distinction we find for even values of $N$
we can choose 
$\big| A\big>$ and $\big| B\big>$ so that
 ${\mathcal P}\big| A \big> = \big| B \big>$ holds.
Therefore ${\mathcal P}\big(\big| A \big> \pm \big| B \big>\big) 
=\pm \big(\big| A \big> \pm \big| B \big>)$
in that case.
Then for even value of $N$ the degeneracy of the ground state
at level-crossing points is not lifted by an electric 
field and no linear Stark effect does appear.

\begin{figure}
\centerline{\psfig{figure=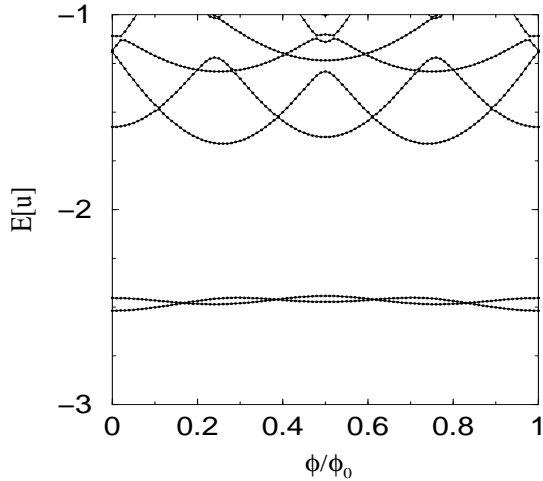,height=6.5cm,width=0.4\textwidth}}
\caption{ The eigenenergies of four quantum rings 
as a function of the threaded magnetic flux per ring $\phi$
 calculated by numerical diagonalization
with the barrier 
potential of $V(\theta)=0.5u\cos(2\theta)$, the interaction of
$U_{|i-j|}(\theta_{i}-\theta_{j})=2u\cos(\theta_{i}-\theta_{j})/|i-j|^{3}$ 
and applied electric field  ${\mathcal E}=0.1u/p_{0}$.  }
\label{n4j2vd05e01ene}
\end{figure}

In Fig. \ref{n4j2vd05e01ene}, 
we show calculated  low energy excitations of 
a system of four rings $(N=4)$ with a potential
 $V(\theta)=0.5u\cos(2\theta)$. 
The dipolar interaction is 
$U_{|i-j|}=J\cos(\theta_{i}-\theta_{j})/|i-j|^{3}$ with  $J=2u$
and the applied electric field is ${\mathcal E}=0.1u/p_{0}$.
Note the clear seperation of the lowest two energy levels
from the remaining part of the spectrum.
The flux dependence of the level excitation energy
is proportional to $\cos(N\pi\phi/\phi_{0})$ as expected.
It is also seen that the degeneracy at the level crossing points
is not lifted by the applied electric field.
This agrees with the general theory discussed above.

It is worth mentioning that the correlated tunneling 
motion of the particles can not be obtained 
if each ring is modelled by a two-level tunneling system.
In that case one has eliminated too many degrees of freedom
in order to describe a correlated particle motion.
Inclusion of higher excited states is necessary 
in order to describe 
the inter-particle interactions during the process of
tunneling which is responsible for the correlated tunneling.

The Hamiltonian $H_{0}$ in Eq. (\ref{hamiltonian}) describes also
an array of Josephson junctions.
In that case $\theta$ is the phase of the superconducting order parameter,
the angular momentum operator goes over into the Cooper-pair
number operator and the kinetic energy is replaced
by the charging energy of a superconducting 
grain\cite{abeles,mclean,simanek,efetov,doniach,review_jja}.
When the superconducting paring is of the conventional form, i.e.,
s-wave, spin singlet paring, the sign of of $J$ in Eq. (\ref{hamiltonian})
is negative.
Therefore no parity effect is expected because
Eq. (\ref{operation_even}) does not hold in that case.
However, an array of $\pi$ junctions should show
a difference between an even and odd number of junctions since
$J>0$ in that case.
This difference refers to the response of the superconducting phase
to a pertubation acting on it.

A different parity effect  which is sometimes called
{\it Leggett's conjecture} 
has already attracted strong attention\cite{leggett}.
In that case, the magnetic field response of $N$ spinless fermions in
a ring show diamagetic (paramagnetic) when $N$ is even (odd)
\cite{leggett,loss}.
In contrast, our parity effect refers to electric field 
response.
In both cases, however, essential roles are played by certain
symmetries in the systems.
While in {\it Leggett's conjecture}
 the antisymmetry of the many-body wavefunctions 
of the fermions are responsible for the even-odd parity effect 
\cite{leggett},
the mirror reflection symmetry in the stack of rings is essential
in the present parity effect.

In summary, we have presented here a theory of stacked 
quantum rings with repulsive inter-ring interactions
in the presence of an applied magnetic field.
Due to the mirror reflection symmetry, the electric 
polarization induced by an applied electric field shows a different
behaviours for even and odd number of rings.
A linear Stark effect at different magnetic fields is expected
for odd number of rings while for even number of rings
the quadratic Stark effect prevails.
In the latter case the polarization is weakly flux dependent
and may show discontinuities at special external magnetic fields.
Nanoscopic semiconductor rings or stacks of ring 
molecules may show the symmetry effects discussed here.
\\

We would like to thank S. Kettemann and A. Bernert for useful and
interesting discussions.

\end{document}